# Independence of the inverse spin Hall effect with the magnetic phase in thin NiCu films

Sara Varotto[1,2], Maxen Cosset-Cheneau[1], Cécile Grezes[1], Yu Fu[1], Patrick Warin[1], Ariel Brenac[1], Jean-François Jacquot [3], Serge Gambarelli[3], Christian Rinaldi[2], Vincent Baltz[1], Jean-Philippe Attané[1], Laurent Vila[1], Paul Noël[1]

[1]Univ. Grenoble Alpes, CEA, CNRS, Grenoble INP, SPINTEC, F-38000 Grenoble, France

[2]Department of Physics, Politecnico di Milano, 20133 Milano, Italy

[3]Univ. Grenoble Alpes, CEA, SYMMES, F-38000 Grenoble, France

**Abstract:**

Large spin Hall angles have been observed in 3$d$ ferromagnets, but their origin, and especially their link with the ferromagnetic order, remain unclear. Here, we investigate the inverse spin Hall effect of $Ni_{60}Cu_{40}$ and $Ni_{50}Cu_{50}$ across their Curie temperature using spin pumping experiments. We evidence that the inverse spin Hall effect in these samples is comparable to that of platinum, and that it is insensitive to the magnetic order. These results points towards a Heisenberg localized model of the transition, and suggest that the large spin Hall effects in 3$d$ ferromagnets can be independent of the magnetic phase.

**Text:**

Heavy metals and dilute alloys with heavy metal impurities offer large spin Hall angles (SHA), as the spin Hall effect (SHE) is intimately related to the spin-orbit coupling [1]. This has motivated a considerable experimental and theoretical effort towards the study of SHE in 5$d$ materials and their alloys [2,3,4,5,6,7,8]. Despite the weaker spin-orbit coupling [9], large SHAs, comparable to that of Pt, can also be obtained in 3$d$ ferromagnetic metals [10,11,12,13,14]. The ferromagnetic material can even have a strong contribution to the spin Hall effect measured in ferromagnet\heavy metal bilayers, through the self-induced SHE [15,16] or the anomalous spin-orbit torque [17]. The spin-orbit coupling is at the origin of specific transport properties in ferromagnets. In particular, its interplay with magnetism gives rise to the anisotropic magnetoresistance [18] and the anomalous Hall effect [19]. However, in ferromagnetic metals, the link between the SHE and the magnetic order is still unclear.

Recent theoretical [20,21] and experimental works [22] showed the existence of two contributions to the SHE in a ferromagnet: a magnetization-independent effect, usually called spin Hall effect, and a magnetization-direction dependent one, known as the anomalous spin Hall effect (ASHE). When the spin polarization of the spin current is aligned with the magnetization the effect is named longitudinal spin Hall effect and is the sum of both terms [23]. Due to the role of the magnetic order, one can expect modifications of the longitudinal spin Hall effect when crossing the Curie temperature ascribed to the symmetry breaking associated to the paramagnetic-ferromagnetic phase transition. Moreover, the modifications of the band structure associated with the phase transition [24,25,26], such as the band shift associated with the vanishing of the exchange splitting [27], could lead to a change in the spin Hall properties. The investigation of the SHE or its reciprocal effect–the inverse spin Hall effect (ISHE) in the paramagnetic state and the inverse longitudinal spin Hall effect (ILSHE) in the ferromagnetic state would therefore provide further insight about the relation between the magnetic order and the spin Hall effect.

The influence of the ferromagnetic phase transition on SHE has been studied in dilute magnetic alloys in a 4d or 5d matrix. The enhancement of the SHA at the Curie temperature in $Ni_{0.09}Pd_{0.91}$ [28] or $Fe_{0.25}Pt_{0.75}$ [29] was attributed to skew scattering on magnetic impurities associated with spin fluctuations [30]. In these alloys, except at the vicinity of the transition temperature, the conversion mechanism is dominated by that of Pt or Pd and the magnetic order plays a little role in the SHE. In another ferromagnetic system, a 4d oxide, $SrRuO_3$ (SRO) the effect of the phase transition on the SHE is reported to be either strong [31] or very weak [32].

In this study, we investigate 3d alloys of NiCu with a stoichiometry dependent Curie temperature [33,34]. The importance of the spin-orbit interaction in this system is known since the first experimental development of the Valet-Fert model [35]. More recently, Keller *et al.* [25] showed that a large SHA could be obtained in NiCu, making it an ideal system to study the effect of the ferromagnetic-paramagnetic transition. We measure the temperature dependence of the inverse spin Hall Effect (ISHE) of NiCu alloys using spin pumping by ferromagnetic resonance (SP-FMR). First, we show that the direct magnetic coupling between the ferromagnetic spin injector and the NiCu has to be avoided, as it generates profound modifications of the dynamic properties of the system below the Curie temperature, which hinder any accurate measurement. Then, when properly decoupled from the spin injector through addition of a Cu spacer layer, we demonstrate large SHE in NiCu, comparable to that of Pt and independent of its magnetic state. More generally, this result evidences that the spin Hall effect in light metal alloys of 3d magnetic elements hold potential for large conversion efficiencies, in both their paramagnetic and ferromagnetic phases.

To perform the spin pumping measurements, we used CoFeB as the spin injector, because its Gilbert damping, its magnetization and its resistivity [36] are nearly temperature independent in the following experimental range, and since its self-induced spin Hall effect is small [16]. Using magnetron sputtering, we grew a 15 nm thick $Co_{53}Fe_{27}B_{20}$ (CFB) layer onto a Si\SiO$_2$ substrate, and then the NiCu alloy layer on top (of different thicknesses in each sample). The NiCu layers were sputtered directly from $Ni_{60}Cu_{40}$ and $Ni_{50}Cu_{50}$ targets. The samples were further capped by a 3 nm aluminum layer to protect them from oxidation. To compare the spin to charge conversion efficiency of NiCu with that of platinum, we also grew a CFB(15)\Pt(15)\Al(3) sample (the numbers in parenthesis indicates the thicknesses in nm), using the same deposition chamber. The samples were then cut into slabs of length $L = 2.4$ mm and width $W = 0.4$ mm.

The SP-FMR measurement were performed in a 3loop-2gap Bruker ER 4118X-MS5 resonator. Following the theory of spin pumping, at the ferromagnetic resonance a spin current is injected from the ferromagnetic layer towards the attached layer [37]. This spin current is then converted into a charge current through ISHE, detected as a voltage in open circuit conditions [2]. As the measured voltage is proportional to both the square of the radiofrequency magnetic field $h_{\text{rf}}$ and the total longitudinal resistance of the sample $R$ [38], we compare these samples on the basis of the normalized spin pumping signal $V/Rh_{\text{rf}}^2$.

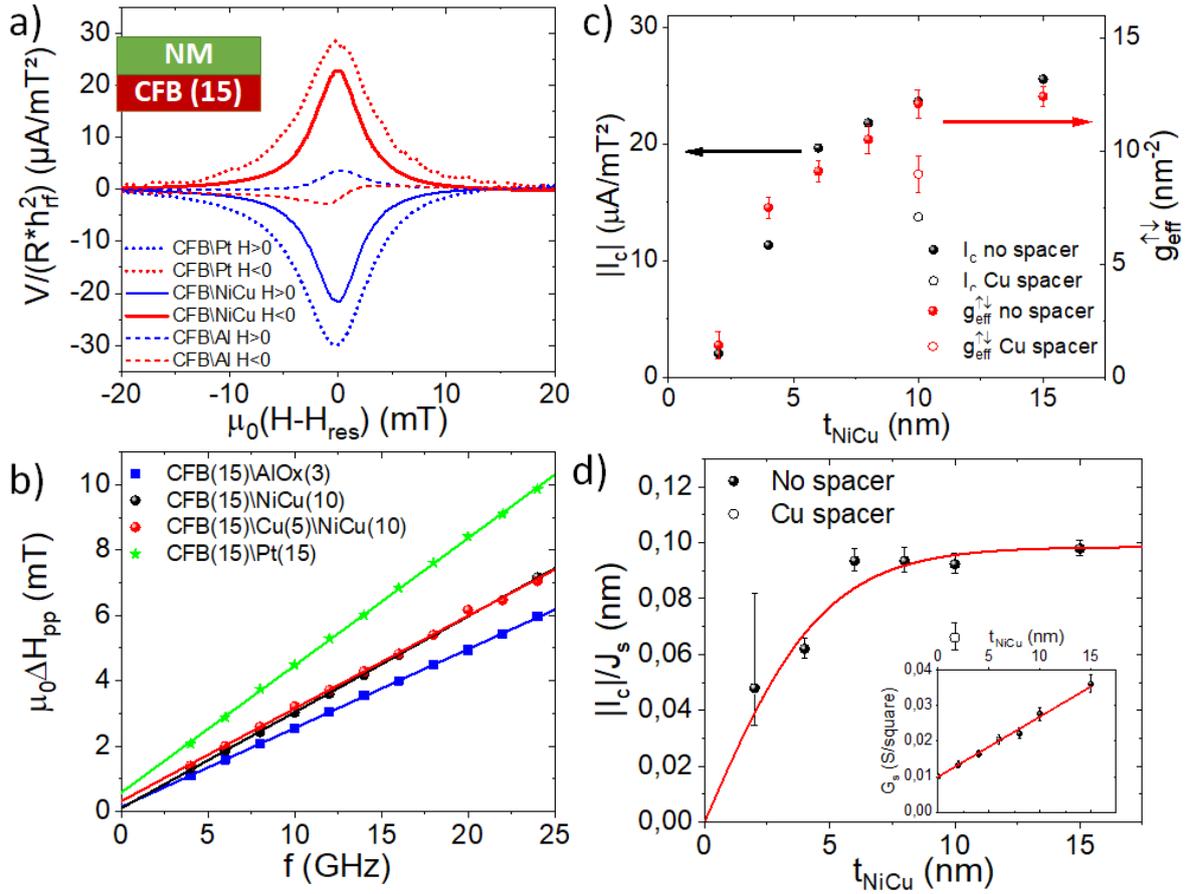

FIG. 1. (a) Spin pumping signals in CFB(15)\Pt(15), CFB(15)\Ni$_{60}$Cu$_{40}$(10) bilayers and in the CFB(15)\Al(3) reference sample. The signals are given in the parallel (*H*>0) and antiparallel (*H*<0) configurations. (b) Broadband measurement of the FMR linewidth of CFB(15)\Al(3), CFB(15)\Ni$_{60}$Cu$_{40}$(10), CFB(15)\Cu(5)\Ni$_{60}$Cu$_{40}$(10) and CFB(15)\Pt(15) samples. (c) Effective spin mixing conductance and absolute value of the charge current production, as a function of the NiCu film thickness $t_{NiCu}$. (d) Thickness dependence of the absolute value of the charge current production divided by the spin current. The fit in red is performed using Eq. 2. Inset shows the thickness dependence of the sheet conductance. The measurements were performed at room temperature, all thicknesses are in nm.

Figure 1(a) shows the spin pumping signal at room temperature obtained at ferromagnetic resonance in CFB(15)\Al(3), CFB(15)\Pt(15) and CFB(15)\Ni$_{60}$Cu$_{40}$(15). The spin pumping signal is dominated by its symmetric part $V_{\text{sym}}$, suggesting that it originates from the ISHE as further confirmed by the out-of-plane angular dependence of the symmetric part of the spin signal, see supplemental material [39]. Remarkably, the spin pumping signal has the same sign and a similar amplitude as in the CFB\Pt sample.

We also performed broadband FMR measurements from 4 to 24 GHZ using a stripline. By applying the Kittel formula $f_{\text{res}} = \frac{\mu_0 \gamma}{2\pi}\sqrt{(M_s + H_{\text{res}} + H_k)(H_{\text{res}} + H_k)}$ with $f_{\text{res}}$ the resonance frequency, $H_{\text{res}}$ the resonance field, $\mu_0$ the vacuum permeability, $\gamma$ the gyromagnetic ratio, $H_k$ the anisotropy field and $M_s$ the effective magnetization of the thin film, we extracted the magnetic properties of the CFB film. The obtained values are similar in all the films and close to previous observations in CFB films of similar stoichiometry, with an effective magnetization of 890 ± 20 kA/m, and a gyromagnetic ratio (*g* factor) of (1.870 ± 0.005)×10$^{11}$ rad.s$^{-1}$·T$^{-1}$ (*g* = 2.13 ± 0.005). We also extracted the Gilbert damping $\alpha$ by fitting the frequency dependence of the peak-to-peak linewidth $\Delta H_{pp}$ using $\mu_0 \Delta H_{pp} = \frac{2}{\sqrt{3}}\frac{2\pi \alpha f}{\gamma} + \mu_0 \Delta H_0$, where $\Delta H_0$ is the inhomogeneous broadening. As can be seen in Fig. 1(b) an enhancement of the damping is observed from (6.64 ± 0.03)×10$^{-3}$ in CFB(15)\Al(3) to (8.05 ± 0.05)×10$^{-3}$ in

CFB(15)\Ni$_{60}$Cu$_{40}$(10) or (1.070 ± 0.006)×10$^{-2}$ in CFB(15)\Pt(15). The damping enhancement, known as extra-damping $\Delta\alpha$, suggests the existence of a spin injection via spin pumping [40], even though it can also be due to other effects such a magnetic proximity [41] and spin memory loss (SML) [42].

We performed cavity spin pumping as well as broadband measurements as function of the thickness of the Ni$_{60}$Cu$_{40}$ layer (from 2 nm to 15 nm), to evaluate both the spin diffusion length $\lambda_s$ and the spin Hall angle $\theta_{SHE}$. The broadband FMR measurement at room temperature evidences that the extracted magnetic properties of the CFB are unaffected by the contact with NiCu, except for the damping that increases with the NiCu thickness saturating at 10 nm (see supplemental material [39]). The absence of a sharp increase of the damping in ultra-thin NiCu layers is not compatible with a large SML or proximity effect. We also grew a sample with a 5 nm insertion of Cu between the CFB and NiCu in order to suppress any magnetic proximity effects [41]. The damping only is slightly reduced from (8.05 ± 0.05)×10$^{-3}$ to (7.75 ± 0.1)×10$^{-3}$ as shown in Fig. 1(b), further evidencing the minor role of the magnetic proximity effect in the damping enhancement. From the broadband FMR measurement we can calculate the effective spin mixing conductance $g_{eff}^{\uparrow\downarrow} = \frac{4\pi M_s t_{FM}}{\gamma\hbar}\Delta\alpha$, with $t_{FM}$ the thickness of the ferromagnetic layer, and $\hbar$ the reduced Planck constant. The obtained values of $g_{eff}^{\uparrow\downarrow}$, as well as the absolute value of the charge current production at resonance $I_c = \frac{V_{sym}(H>0) - V_{sym}(H<0)}{2Rh_{rf}^2}$, are plotted in Fig. 1(c) with and without the Cu spacer. The small contribution of CFB to the spin pumping signal was separately determined on CFB(15)\Al(3) layer (Figure 1(a)) and consequently subtracted. The spin mixing conductance and spin pumping signal saturate at 10 nm with a value of $12.4 \pm 0.4$ nm$^{-2}$ and 25 µA/mT$^2$, respectively. Owing to the apparent minor role of magnetic proximity effect and SML in the CFB\NiCu bilayers, we evaluate the spin current using the classical spin pumping model [37,38]:

$$J_s = \frac{g_{eff}^{\uparrow\downarrow}\gamma^2\hbar h_{rf}^2}{8\pi\alpha^2}\left(\frac{\mu_0 M_s\gamma + \sqrt{(\mu_0 M_s\gamma)^2 + 4\omega^2}}{(\mu_0 M_s\gamma)^2 + 4\omega^2}\right)\left(\frac{2e}{\hbar}\right). \qquad (1)$$

The resistivity $\rho$ of the NiCu layer is of $60 \pm 3$ µΩ · cm, and independent on the thickness as shown in the inset of Fig. 1(d) in line with the short mean free path of less than 1 nm in similar alloys [43]. We can thus assume a constant spin diffusion length and spin Hall angle in the fit. The charge current as a function of the NiCu thickness is then given by:

$$I_c(t) = W\theta_{SHE}\lambda_s J_s \tanh\left(\frac{t_{NiCu}}{2\lambda_s}\right). \qquad (2)$$

Combining Eqs. (1) and (2) it is possible to extract $\lambda_s$ and $\theta_{SHE}$ from $I_c/J_s$. From the fit in Fig. 1(d) we obtained $\theta_{SHE} = 4.1^{+0.6}_{-0.4}\%$. In contrast to Ref. [25], we did not observe any remarkable spin memory loss or magnetic proximity effect, which might explain why our estimated SHA is much lower. The value of $I_c/J_s$ proportional to $\theta_{SHE}\lambda_s$ is 32 % smaller with respect to the case of direct contact. This is likely since we did not account for the possible SML at the CoFeB\Cu interface, which is typically of the order of 20 to 30 % in CoFe\Cu [44] and Co\Cu [45].

In order to investigate the effect of the NiCu phase transition, we performed a temperature dependence of the spin pumping with and without Cu insertion in samples with 10 nm thick NiCu.

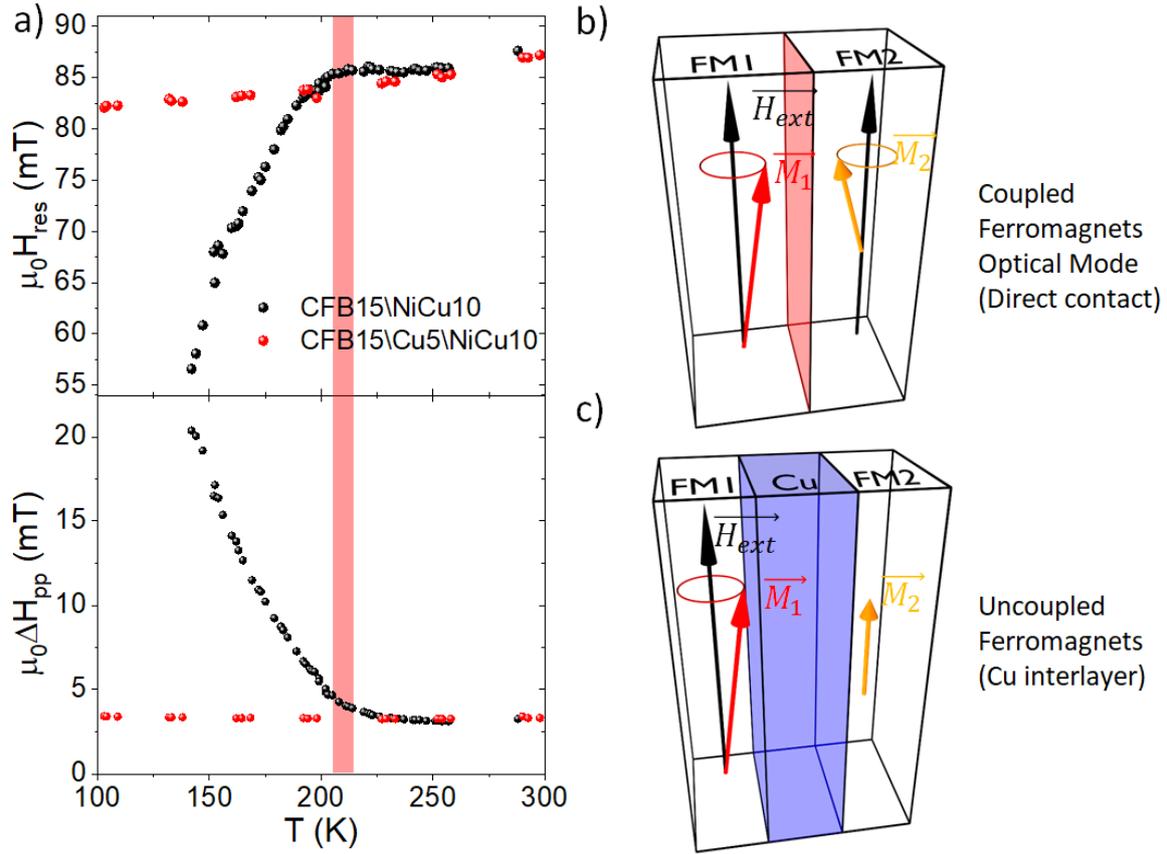

FIG. 2. (a) Resonance field and linewidth of CFB as a function of the temperature in CFB(15)\Ni$_{60}$Cu$_{40}$(10) and CFB(15)\Cu(5)\Ni$_{60}$Cu$_{40}$(10). The red shaded area depicts the Curie temperature $T_c$ (210 ± 5 K) of Ni$_{60}$Cu$_{40}$. (b) In direct contact at resonance both magnetization vectors of FM1 (CFB) and FM2 (NiCu) precess out of phase for the optical mode. (c) When the ferromagnets are decoupled using a thick enough Cu interlayer only FM1, here CFB, is excited at its resonance.

As can be seen in Fig. 2(a) the resonant properties of both samples are similar at room temperature, but their behaviors at low temperature are strikingly different. In absence of the Cu interlayer, the resonance field decreases below 210 K, while the linewidth increases. These effects are not observed in presence of the 5 nm Cu spacer. This change occurs around the expected paramagnetic to ferromagnetic transition in Ni$_{60}$Cu$_{40}$ [33,34] whose Curie temperature is $T_c$ = 210 ± 5K as evidenced by AHE and resistivity measurements (see supplemental material [39]). The value of $T_c$ is not affected by the deposition of Cu or by the proximity with a ferromagnet, contrary to previous observations in spin valves [46].

When CoFeB is in direct contact with NiCu and below $T_c$, an exchange coupling between the two layers arises, so that the magnetization vectors of the two ferromagnets precess together either in phase (acoustical mode) or out of phase (optical mode). The spin transport can be of magnonic nature and is not limited to electrons [47], leading to strong modifications of the resonance field and linewidth [47,48,49]. The lowering of the resonance field in Fig. 2(a) is associated to the optical mode (out of phase precession) depicted in Fig. 2(b). Such a decrease becomes larger when the temperature is lowered, due to both the increased magnetization of NiCu and the strengthening of the exchange coupling. With the 5 nm Cu spacer however, the exchange coupling is suppressed (see supplemental material [39]), the electronic transport prevails, and there is no noticeable change of the resonance field and linewidth around the Curie temperature. As the two magnetic layers are not coupled, when

CFB is at resonance magnetization of NiCu is out of resonance (i.e. static) as depicted in Fig. 2(c), so that the dynamical properties remain unaffected [50].

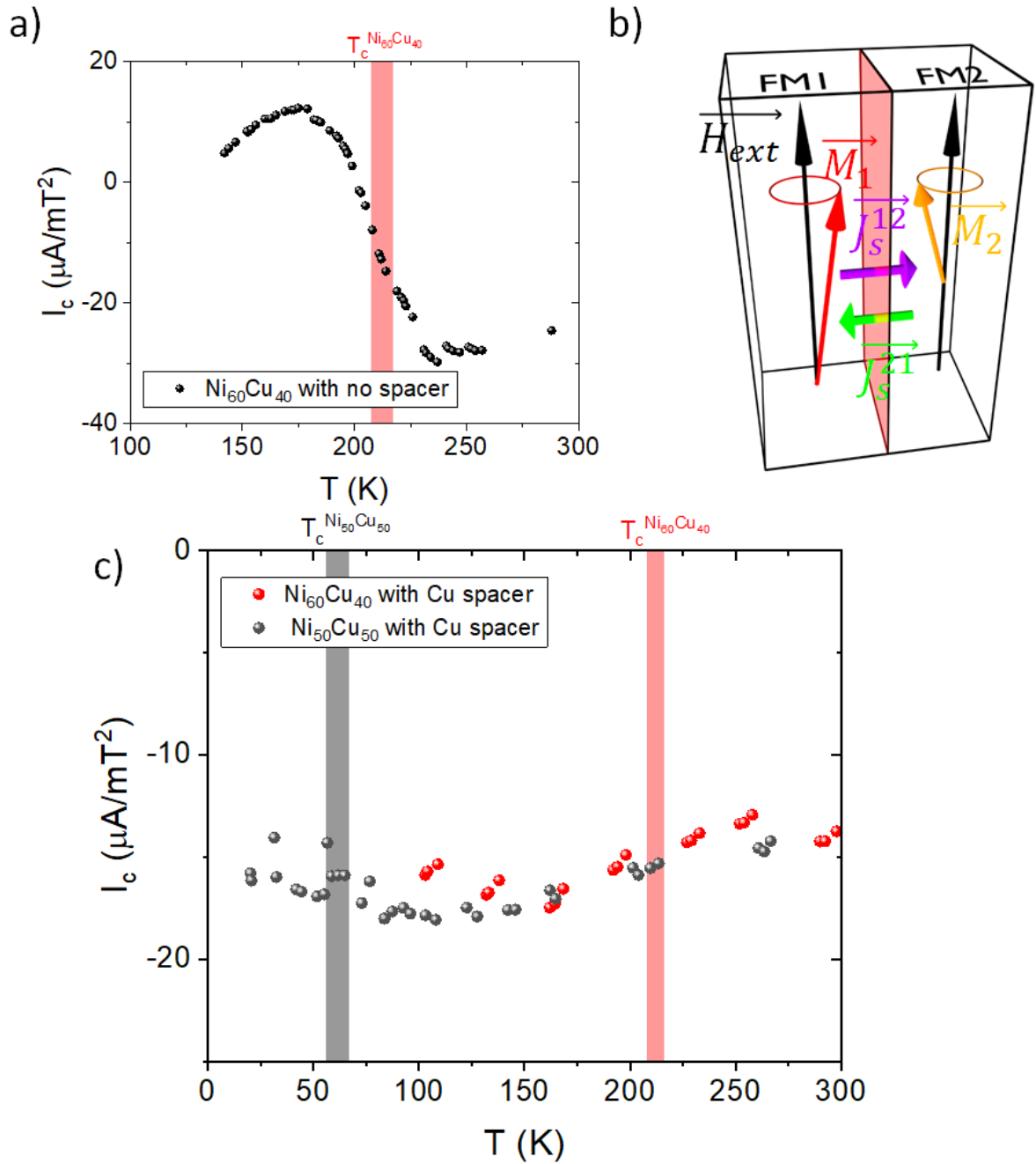

FIG. 3. (a) Spin pumping signal in a coupled bilayer without Cu spacer. In red the $T_c$ of $Ni_{60}Cu_{40}$. (b) Schematic representation of the spin pumping process below $T_c$ (optical mode) for the direct contact. Spin currents are injected from FM1 to FM2 ($J_s^{12}$) and from FM2 to FM1 ($J_s^{21}$) with a 180° phase shift. (c) ISHE signal as a function of temperature in two CFB(15)\Cu(5)\NiCu(10) samples with $Ni_{60}Cu_{40}$ and $Ni_{50}Cu_{50}$ with different stoichiometries and Curie temperatures ($T_c^{Ni60Cu40}$ shown in red and $T_c^{Ni50Cu50}$ in grey).

The charge current production at resonance $I_c$ as a function of the temperature was also measured with and without Cu insertion. As can be seen in Fig. 3(a). in the exchange-coupled system without the Cu insertion $I_c$ is negative above the Curie temperature but when lowering the temperature, the current production is abruptly modified around $T_c$, and even changes sign below 200 K. Without further measurements, one might conclude that the SHE is strongly affected by the ferromagnetic transition, and changes sign on a very narrow temperature range. However, in the case of the direct contact, the strong modification of the resonance properties across the magnetic phase transition makes it impossible to accurately evaluate the injected spin current at resonance below the Curie temperature using the spin pumping model. Moreover, the CoFeB and NiCu layers are coupled and both precessing at resonance, so that a spin current is injected from CFB towards NiCu and from NiCu towards CFB as depicted in Fig. 3(b). The exact spin accumulation profile cannot be evaluated, and a non-negligible ISHE signal from CFB could start contributing significantly [13]. For these reasons, to properly evaluate the ISHE in NiCu below the Curie temperature, it is necessary to perform the same measurement in the CFB\Cu\NiCu trilayer and verify that such sign change is not due to spurious effects related to the exchange coupling. In this configuration below $T_c$ the magnetization of NiCu and the injected spins are parallel so that the measured signal is the inverse longitudinal spin Hall effect. In the decoupled system, $I_c$ is constant and does not change sign at $T_c$, as visible in Fig. 3(c). The absence of any noticeable effect is also observed for another stoichiometry $Ni_{50}Cu_{50}$ ($T_c$ = 60 K). In these trilayers the spin signal as well as the resonant properties of the CFB layer are constant with temperature: the same spin current injection and charge current production are observed in both the paramagnetic and the ferromagnetic states. Therefore, the large ISHE in NiCu alloys in the paramagnetic phase is equal to the inverse longitudinal spin Hall effect in the ferromagnetic phase and is thus unrelated to the magnetic order.

We observed that the spin Hall angle is positive while the anomalous Hall angle is negative (see supplemental material Information [39]), in agreement with the calculations of the intrinsic effects in ferromagnetic Ni [17,20,21]. This suggests that the two mechanisms in NiCu are dominated by the intrinsic contribution. It also confirms previous observations by Omori *et al.* that the anomalous and spin Hall effect can be of opposite sign [14]. For both $Ni_{60}Cu_{40}$ and $Ni_{50}Cu_{50}$, the figure of merit of the spin to charge current conversion $\theta_{SHE}\lambda_s$ is of 0.1 nm independently on the stoichiometry, as expected for alloys [8]. This value is comparable to that of Pt (0.18 nm) [42,51], thereby allowing for efficient detection of spin currents.

The large SHA in paramagnetic NiCu was previously explained by similarities between the calculated band structure of paramagnetic NiCu, paramagnetic nickel and platinum around the Fermi level [25]. Nonetheless the calculated band structure of ferromagnetic nickel in Ref. [25] does not resemble that of Pt. This is not compatible with our observation of a large magnetic phase independent ISHE in NiCu. On the contrary our observations suggest that the band structure is not particularly affected by the phase transition, or that these possible changes do not reflect in a modification of the spin Hall effect properties. The modifications of the band structure occurring at the ferromagnetic transition and in particular the shift of the bands associated with the collapse of the exchange splitting are still debated with conflicting results for nickel [27, 52 , 53 ]. The ferromagnetic-paramagnetic transition is characterized by the loss of the long-range ferromagnetic order but is not necessarily associated with the collapse of exchange splitting that occurs at higher temperature as observed in $SrRuO_3$ or Fe [54,55]. These results evidence that the itinerant Stoner model of ferromagnetism does not describe well the ferromagnetic-paramagnetic phase transition in NiCu. The Heisenberg localized model seems to be more adequate. This model is based on the existence of local fluctuating magnetic moments above the Curie temperature which is likely to be the case of NiCu, where local moments associated to the nearest neighbor environment of nickel atoms are at the origin of magnetism [56,57]. In that scenario, the disappearance of the magnetization above $T_c$ is not due to a shift of the majority and minority band but to band-mirroring [26]. Thus, our experimental results could be explained by similar band structure of NiCu in the paramagnetic and ferromagnetic phase.

In summary, using spin pumping FMR in CFB\Cu\NiCu with two different stoichiometries, we measured a large positive spin Hall angle of NiCu comparable to the one of Pt, which is insensitive to the magnetic phase when magnetization and injected spins are aligned. This insensitivity suggests that the effect of the ferromagnetic-paramagnetic transition on the SHE has to be understood, at least in NiCu, using a localized Heisenberg model rather than the itinerant electron Stoner model. We emphasized the importance of avoiding direct contact to properly measure the ISHE in a ferromagnet for resonance measurements. Not doing so leads to drastic changes in the dynamic response that are hard to estimate. This might explain the discrepancy between measurements of the spin Hall effect in the same material but in contact with different ferromagnets [29,31]. Moreover, our results show that the ferromagnetic order does not necessarily play a key role in the large spin Hall effect of ferromagnets. A large variety of alloys composed of Ni, Co or Fe in the ferromagnetic or paramagnetic phase could therefore be explored as spin current generators and detectors, extending the number of possible light metal systems with a large SHE.

**Acknowledgments:**
We acknowledge the financial support by ANR French National Research Agency Toprise (ANR-16-CE24-0017), ANR French National Research Agency OISO (ANR-17-CE24-0026), Laboratoire d'excellence LANEF (ANR-10-LABX-51-01) and the Fondazione Cariplo and Regione Lombardia, grant No. 2017-1622 (ECOS). We are grateful to the EPR facilities available at the national TGE RPE facilities (IR 3443). We thank Christian Lombard and Vincent Maurel for their help and advices on the FMR measurement setup. Interesting discussions with Yasuhiro Niimi are gratefully acknowledged.